# Spontaneous photo-generated carrier separation of SnO/BiOX (X=Cl, Br, I) bilayer under visible light irradiation for water splitting

Yanyu Liu[1], Peng Lv[1], Wei Zhou[2*], Jiawang Hong[1*]

[1] School of Aerospace Engineering, Beijing Institute of Technology, Beijing 100081, China

[2] *Department of Applied Physics, Tianjin Key Laboratory of Low Dimensional Materials Physics and Preparing Technology, Faculty of Science, Tianjin University, Tianjin 300072, P.R. China*

**Abstract**

Alloying in 2D materials plays a more and more important role due to wide range bandgap tunability and integrating the advantages of HER and OER. Here, the novel bilayers of SnO/BiOX (X= Cl, Br and I) bilayer are constructed to integrate the advantages of narrow bandgap and separating photo-generated carriers. The bandgap of the bilayers can be tuned from 1.09 to 1.84 eV, remarkably improving the utilization of solar energy. The large difference in effective masses and built-in electric field effectively hamper the fast recombination of photo-generated carries, which highly enhances the photocatalytic efficiency. Besides that, the type-II band alignment guarantee the two half reactions could occur at different surfaces. Moreover, the optical absorption (the strong transition between band edges and high joint density of states) and band-edge level further confirm the SnO/BiOX (X= Cl and Br) bilayer is a promising candidate for overall water-splitting.

———————————————————————

*Corresponding authors: hongjw@bit.edu.cn
　　　　　　　　　　　weizhou@tju.edu.cn

**INTRODUCTION**

Inspired by photosynthesis, artificial photosynthesis is attracting ever-growing attention[1]. Solar cell and photocatalytic water-splitting using semiconductor yields the conversion of the solar energy into electric energy and chemical fuel, respectively. In addition to a free and unlimited supply of solar energy and water, the fascination is also from the features of zero pollution, portable fuels and freedom of weather restrictions. Despite the above aforementioned advantages, the large-scale commercial application is still hampered by the following reasons: unable to utilize visible light (the optimal band gap for water splitting and solar cell is 1.23 eV[2-5] and 1.50 eV[6]), high recombination of photo-generated electron/hole pairs and unsuitable band-edge positions. In brief, the insufficient effective photo-generated carriers with suitable energy level are responsible for their poor conversion efficiency in catalysis[7]. To gain essential effective photo-generated carriers, considerable efforts have been made[4], such as ion doping[8-10] and construction of bilayer[11-15] to increase light harvesting, in addition of sacrificial reagents to enhance the photo-generated electron/hole separation.

However, there are several drawbacks to the above strategies[4, 8]. For example, the deep impurity states induced by doping may act as recombination centers for photo-generated electrons and holes, more sacrificial reagents is required to sustain water splitting[16]. Therefore, constructing a type II bilayer plays a more and more significant role in improving photocatalysts and has been extensively studied[14, 17-20] due to its effective charge separation feature. Besides the separation of the photo-generated carriers, another advantage of type-II band bilayer is to harvest rich visible light by the synergic absorption of two semiconductors with cross band gap and drive the photo-generated electrons and holes to move in opposite directions by built-in electric field[11].

Based on previous experimental and theoretical reports, the solar conversion efficiency of type-II band bilayers outperforms their counterpart[18, 21-23]. For example, the current density of α-Fe$_2$O$_3$/Zn$_{0.4}$Cd$_{0.6}$S heterostructure increases 3~4 times, compared with the pristine α-Fe$_2$O$_3$ or Zn$_{0.4}$Cd$_{0.6}$S[24]. The H$_2$ evolution rate reaches 49.80 mmol g$^{-1}$h$^{-1}$, far exceeding 0.35 mmol g$^{-1}$h$^{-1}$ of CdS nanorods alone by a factor of 130.26[25]. Some theoretical works predict the spatial separation of photo-generated electron-hole pairs, resulting in two half-reactions of water splitting occurs at the different surface of materils[12-13, 17]. For example, ZnS/SnO bilayer, BiOBr/BiOI bilayer are predicted that the desirable band alignment and band gap render them a viable candidate for optoelectronic applications, in particular for solar water splitting[13, 26]. However, despite the excellent achievement in laboratory for decades, it is still scarce for low-cost, high efficiency and tunable band gap materials for photo-conversion applications.

Herein, we present the interface engineering of novel SnO/BiOX (X= Cl, Br and I) bilayers based on the type-II band alignment and good lattice match. According to our results, the narrow band gap, large potential drop, the spatial separation of photo-generated electrons and holes and the dramatic difference of effective mass of electron render SnO/BiOX bilayers promising optoelectronic materials. The band alignment confirms that the SnO/BiOCl and SnO/BiOBr bilayers possess the desirable band-edge positions for the redox water splitting capability. Furthermore, based on Shockley-Quisser limit, the band gap of 1.47 eV also makes the SnO/BiOBr bilayer as an efficient solar conversion material. These provide more and better candidate materials for photochemical and optoelectronic applications.

**COMPUTATIONAL DETAILS**

All calculations for SnO/BiOX hybrid structures are carried out based on density functional theory (DFT) with the projector-augmented-wave (PAW) pseudopotentials[27] implemented in the Vienna *ab initio* simulation package (VASP)[28-29]. The generalized gradient approximation (GGA) of the Perdew-Burke-Ernzerhof (PBE) scheme to depict the exchange and correlation potential[27]. The empirical correction method proposed by Grimme (DFT-D2)[30] is used to describe the vdW interactions. A kinetic energy cutoff in the plane-wave expansion is 450 eV. To ensure that the interactions between the slabs along the vertical direction are negligible, the vacuum space of 20 Å is set. The 2D Brillouin zone is sampled with 7×7×1 k-points within the Monkhorst-Pack scheme for all the calculations. Both lattice constants and atomic positions are relaxed until the convergence criterial of energy and force are less than $10^{-5}$ eV and 0.01 eV Å$^{-1}$, respectively. Considering the band gap underestimation of PBE functional, the Heyd-Scuseria-Ernzerhof (HSE06) hybrid functional are chosen for band-gap and electronic properties calculations[31-32].

To reveal the inherit nature of charge transfer and separation in the SnO/BiOX bilayers, the 3D spatial charge density difference is calculated. The 3D special charge density difference provides a quantitative analysis for the interfacial charge redistribution, which is defined as:

$$\Delta\rho = \rho_{heterostructure} - \rho_{SnO} - \rho_{BiOX} \qquad (1)$$

where $\rho_{heterostructure}$ is the charge density of the SnO/BiOX bilayer, while $\rho_{SnO}$ and $\rho_{BiOX}$ are the charge density of isolated SnO and BiOX monolayer, respectively.

As one of the criteria for the optoelectronics applications, the optical absorption is also studied. According to the Fermi Golden rule, the imaginary part of the dielectric function is first calculated by the following formula[33]：

$$\varepsilon_2(\omega) = \frac{2\pi}{\hbar} \int \langle v|\hat{H}|c\rangle^2 \frac{2}{8\pi^3} \delta(E_c(\vec{k}) - E_v(\vec{k}) - \hbar\omega) \, d^3k \tag{2}$$

where $\langle v|\hat{H}|c\rangle$ is the transition matrix from states in the valence band to that in the conduction band. The second term is the joint density of states (JDOS) at energy $\hbar\omega$. Then, the real part $\varepsilon_1(\omega)$ of the dielectric function could be obtained by the Kramer-Kronig transformation[34]:

$$\varepsilon_1(\omega) = 1 + \frac{2}{\pi} P \int_0^\infty \frac{\omega' \varepsilon_2(\omega') d\omega'}{(\omega'^2 - \omega^2)} \tag{3}$$

Finally, the expressions for the absorption coefficient $\alpha(\omega)$ now follow immediately[35]:

$$\alpha(\omega) = \sqrt{2}\omega \left[ \sqrt{\varepsilon_1^2(\omega) + \varepsilon_2^2(\omega)} - \varepsilon_1(\omega) \right]^{1/2} \tag{4}$$

Therefore, the optical absorption of a semiconductor at photonic energy $\hbar\omega$ is directly correlated with the transition matrix and JDOS.

**RESULTS AND DISCUSSION**

**Crystal structure for bilayer**

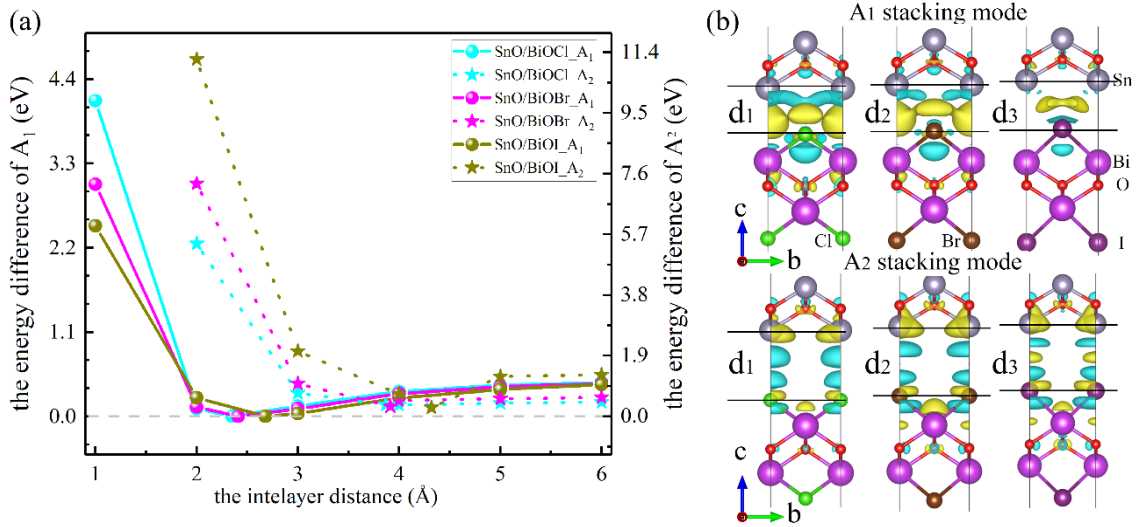

Fig. 1 The energy difference as a function of interlayer distance, d (a) and charge density difference (b) of the SnO/BiOX bilayer in $A_1$ (a scale unit of $4\times10^{-4}$ e/Å$^3$) and $A_2$ (a scale unit of $8\times10^{-5}$ e/Å$^3$) stacking patterns. The cyan and yellow areas indicate electron depletion and accumulation, respectively.

To simulate the SnO/BiOX hybrid structures, a 1×1 supercell of monolayer SnO on top of a 1×1 supercell of BiOX is built. Since the lattice mismatch between SnO and BiOX within 2%, the hybrid structures are feasibly formed. The SnO/BiOX hybrid structure consists of two stacking patterns by changing the relative position of Sn atoms on the BiOX layer: the Sn atoms of SnO are right above the Bi atoms (hereafter remarked as $A_1$) or X atoms of BiOX (hereafter remarked as $A_2$). In order to obtain the optimal interlayer distance, d, between SnO and BiOX, the energy difference as a function of d is plotted in Fig. 1(a). Obviously, the $A_1$ stacking pattern is more stable than $A_2$ stacking configuration due to the smaller energy for the same constituent. And the optimized vertical interlayer distances of SnO/BiOX bilayer with $A_1$ pattern of $d_1$, $d_2$ and $d_3$ are much shorter than that of $A_2$ pattern (2.38, 2.44 and 2.68 Å for $A_1$ pattern vs. 3.87, 3.91, 5.63 Å for $A_2$ pattern).

The large difference in the interlayer distance could be explained by the bonding charge density of the SnO/BiOX bilayer, as presented in Fig. 1(b). It can be seen that the interaction for the $A_1$ stacking pattern is much stronger than that of $A_2$ stacking pattern, giving rise to the shorter interlayer distance in the $A_1$ structures. In addition, with increasing atomic numbers, the vdW equilibrium separation between SnO and BiOX layer become larger. This is related to the decreased electronegative with increasing the atomic number, which results in the charge-attractive forces decreases between the Sn and X atoms of the constituent layers at the coupling interface. Consequently, the distance of the SnO and BiOX layer becomes larger with increasing atomic numbers.

To evaluate the stability of SnO/BiOX bilayer, the binding energies ($E_b$) are calculated by $E_\mathrm{b} = E_\mathrm{SnO/BiOX} - E_\mathrm{SnO} - E_\mathrm{BiOX}$, where $E_\mathrm{SnO}$, $E_\mathrm{BiOX}$ and $E_\mathrm{SnO/BiOX}$ are the total energies of isolated SnO, BiOX, and SnO/BiOX hybrid structures, respectively. The calculated binding energies have been summarized in Table I. It clearly shows that the Sn atoms of SnO is energetically favorable on top of Bi atoms, due to the smaller binding energy. Therefore, in the following discussions, we will only focus on the A₁ structure. The interlayer distances of SnO/BiOX bilayer with A₁ stacking pattern are much shorter than the exemplary bilayer of graphene/graphene (3.34 Å)[36] and BN/BN (3.33 Å)[37]. Hence, the coupling strength of SnO/BiOX junction is stronger than the cases of graphene/graphene and BN/BN.

Table I Calculated binding energy of SnO/BiOX bilayers with different stacking models, in eV.

|  | SnO/BiOCl | SnO/BiOBr | SnO/BiOI |
|---|---|---|---|
| $E_b$ (A₁) | -0.54 | -0.36 | -0.29 |
| $E_b$ (A₂) | -0.09 | -0.07 | -0.01 |

**Type-II band alignment and optical absorption properties**

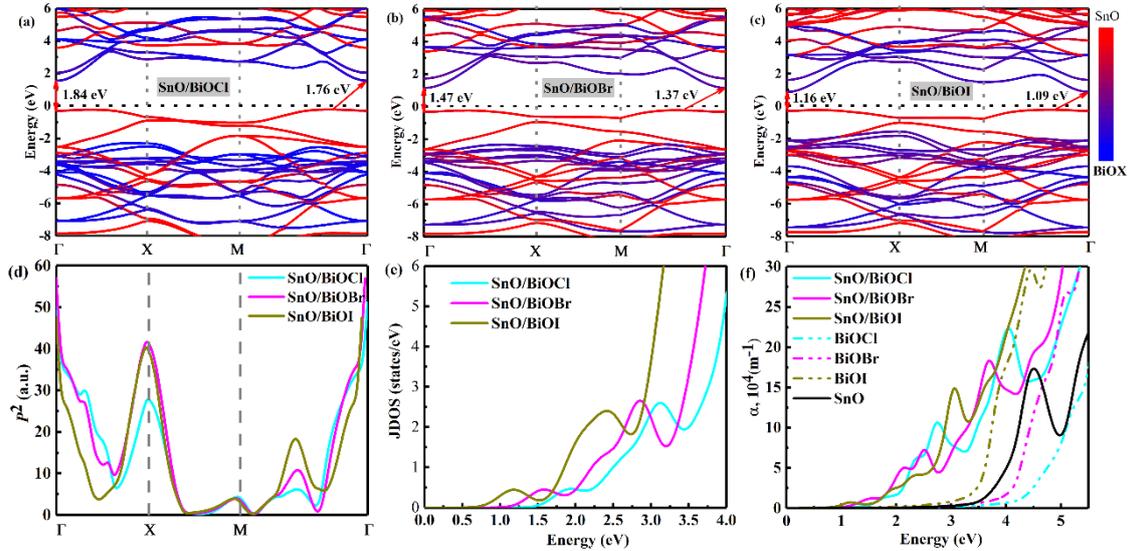

Fig. 2 The band structures (a)-(c), transition matrix elements (d), JDOS (e) and optical spectra (f) for SnO/BiOX bilayers.

The band structures along the high-symmetry directions of the Brillouin zone (BZ) for the SnO/BiOX junction are plotted in Fig. 2(a)-(c). The calculated band gaps for the SnO/BiOX junctions with the isolated SnO and BiOX sheets are summarized in Table II. The bandgap of the bilayers can be tuned from 1.09 to 1.84 eV, which remarkably improves the utilization of solar energy. Besides that, the energy band gaps of the SnO/BiOX junction significantly reduce after stacking. And the difference between direct and indirect bandgap values is within 0.1 eV. Therefore, the band gaps exhibit nearly direct bandgaps for all SnO/BiOX bilayers.

Table II The direct (D) and indirect (I) band gap of SnO/BiOX junction and the isolated SnO and BiOX sheets, in eV.

|  | BiOCl | BiOBr | BiOI | SnO |
| --- | --- | --- | --- | --- |
| Isolated sheet | $4.62^D/3.77^I$ | $3.91^D/3.37^I$ | $2.75^D/2.31^I$ | $3.90^D/3.86^I$ |
| SnO/BiOX | $1.84^D/1.76^I$ | $1.47^D/1.37^I$ | $1.16^D/1.09^I$ | - |

Although the direct band gap is a desirable nature for optoelectronic applications, the dipole transition matrix elements and the JDOS determine the optical absorption of a semiconductor at a specified photon energy, based on eqs. (2)-(4). Additionally, parity-forbidden transitions between conduction and valence band edges are particularly undesirable for solar absorber because of the inefficient absorption of photons with the energy close to the band gap value[38-39]. The sum of the squares of the dipole transition matrix elements $P^2$ can tell the high transition probabilities between the topmost valence and the lowest conduction band at various k points. Therefore, we calculate the $P^2$ and JDOS to reveal the optical absorption capability of SnO/BiOX junctions, as shown in Fig. 2(d)-(e). Obviously, there is no parity-forbidden transition for all SnO/BiOX junctions due to the strong transition between conduction and valence band edges, and the rather flat top valence band results in a high JDOS. These two features usually yield SnO/BiOX junctions a high solar conversion efficiency. In order to verify this speculation, we calculate the optical absorption coefficient ($\alpha$) of the SnO/BiOX bilayer with its isolated SnO and BiOX sheets with isolated SnO and BiOX sheets, with results presented in Fig.

1(f). From Fig. 1(f), the absorption edges noticeable red-shift occurs after stacking, which is in good consistence with the narrow band gap of the SnO/BiOX junctions. Thus, the solar conversion efficiency of the SnO/BiOX bilayer are much higher than their counterpart isolated sheet.

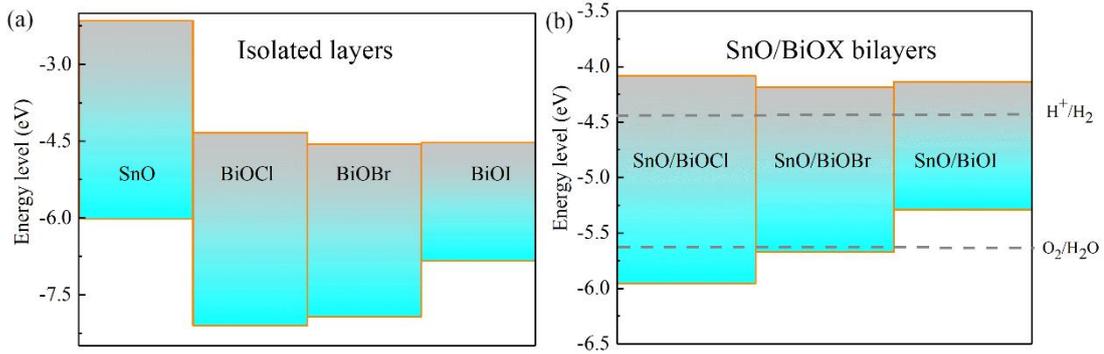

Fig. 3 Diagrams of the band edge positions of isolated layers (a) and SnO/BiOX bilayers (b), the top and bottom of the corresponding rectangles signify the CBM and VBM values, respectively.

To reveal the possible origin of the small band gap, the band alignments are constructed referenced to the vacuum level[40] for the SnO/BiOX bilayer in the pre-contact and contact states (see Fig. 3). Before the formation of bilayer, the CBM of SnO is more negative than that of BiOX, whereas the VBM of BiOX is more positive than that of SnO. Upon SnO and BiOX stacking, the upshift of VBM of BiOX is much larger than that of SnO, whereas the position of CBM of BiOX has no obvious change. As a result, the SnO/BiOX junction demonstrates a small band gap.

As well known, one of the critical requirements in evaluating the photocatalytic ability of semiconductor photocatalyst for water splitting is the band-edge positions with respect to water redox levels. Therefore, the band-edge positions of the SnO/BiOX bilayer are studied with respect to water redox levels, as depicted in Fig. 3(b). Apparently, the CBM of the SnO/BiOX bilayer is higher than the standard reduction potential $H^+/H_2$ (−4.44 eV), while only the VBM of SnO/BiOCl and SnO/BiOBr is lower than the oxidation potential for O$_2$/H$_2$O (−5.67 eV).That is to say, the H$^+$/H$_2$ half reaction well proceeds with the CBM electrons provided by the BiOCl or BiOBr layer constituent, the O$_2$/H$_2$O half-reaction could take place on the SnO layer constituent, as shown in Fig. 3(b). As for the case of SnO/BiOI bilayer, only the reduction capability is retained. Besides, the band gap of the SnO/BiOBr bilayer is about 1.47 eV, indicating it is a potential candidate for solar cell application based on the Shockley-Quisser limit.

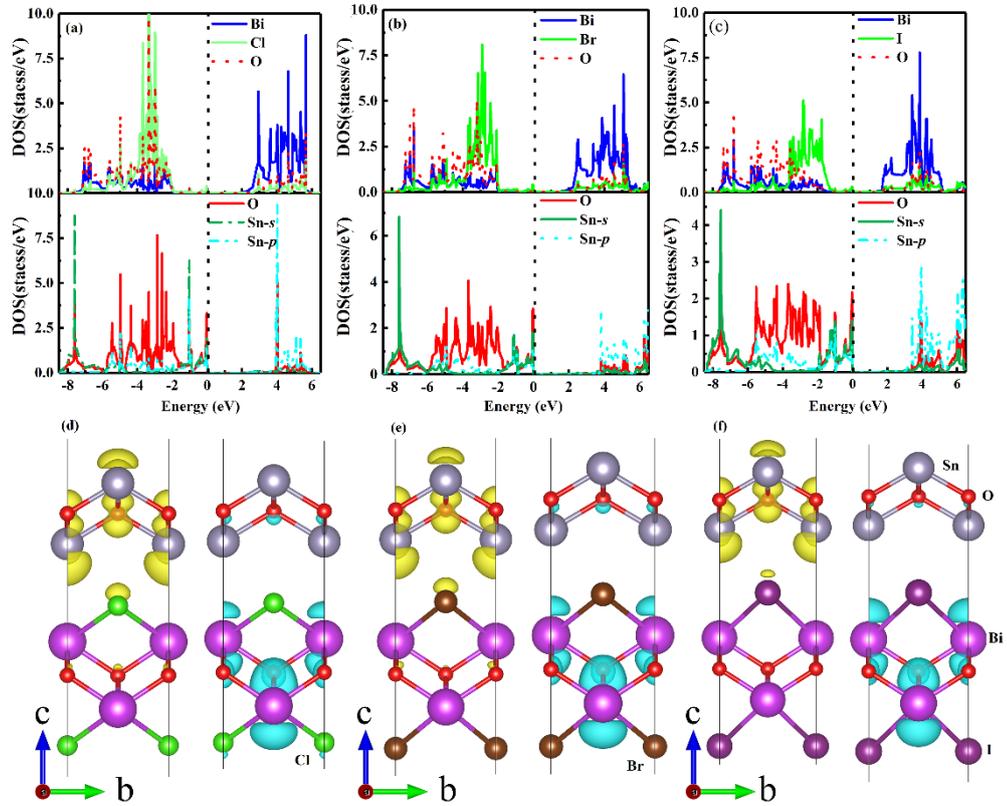

Fig. 4 The DOS and projected charge density of CBM and VBM (a scale unit of 8×10$^{-3}$ e/Å$^3$) for SnO/BiOCl (a) and (d), SnO/BiOBr (b) and (e) and SnO/BiOI (c) and (f). The yellow and cyan areas represent projected charge density of VBM and CBM, respectively.

From the band structures of SnO/BiOX junctions, it could be observed that the lower conduction band and upper valence band are mainly dominated by the electronic states of BiOX and SnO, respectively. Therefore, this family of hybrid structure is type II semiconductor, namely, with both CBM and VBM of BiOX below the corresponding

SnO counterparts. The detailed contributions of VBM and CBM could be analyzed by the projected density of states (DOS) of SnO/BiOX junctions, (see Fig. 4(a)-(c)). The occupied-electron states at VBM are mainly derived by the O and Sn atoms of the SnO layer, while the unoccupied-electron states at CBM are mainly contributed by the Bi atoms the BiOX layer, with also small contribution from O and X atoms. Obviously, the different layer components constitute the top and bottom surface of the SnO/BiOX bilayers. The distinguished electronic structure nature could also be seen by the partial charge densities at CBM and VBM (see Fig. 4(d)-(f)). The partial charge densities visually reflect that the photo-generated holes at VBM are primarily produced on the SnO layer component, while the photo-generated electrons of CBM mainly distribute on the BiOX layer component. The spatial separation of photo-induced carriers also takes place in case of BiOBr/BiOI and g-ZnO/MoS$_2$, which greatly prolongs the life of photo-induced carriers.

**Hard recombination of photo-generated electron-hole pair**

Table III Calculated the effective masses at the CBM ($m_e$) and VBM ($m_h$) for the SnO/BiOX bilayers, in $m_0$.

|  | $m_h$ | $m_e$ |
|---|---|---|
| SnO/BiOCl | 2.01 | 0.29 |
| SnO/BiOBr | 2.09 | 0.26 |
| SnO/BiOI | 1.38 | 0.22 |

Besides the spatial separation of photo-generated electrons-hole pairs, the large difference in electron/hole mobility and directional separation also hold the key to the improvement of photo-conversion efficiency. Therefore, we estimate the effective masses of carriers at the CBM ($m_e$) and VBM ($m_h$) and the planar average electrostatic potential for the SnO/BiOX bilayers. The effective masses of carriers for SnO/BiOX can be calculated by fitting the band edge with the following formula:

$$m^* = \hbar^2 \left(\frac{d^2 E(k)}{dk^2}\right)^{-1} \quad (5)$$

where $m^*$, $\hbar$, k and E($k$) represent the effective mass of the carrier, Planck constant, the wave vector and eigenvalues of energy band, respectively. The calculated effective masses have been summarized in Table III. Obviously, the $m_e$ and the $m_h$ indeed present a significant difference in the bilayer. Based on the reciprocal relationship between effective masses and mobility, the migration speed of electron is faster than that of holes by about 6~8 times. It means the carrier separation could be realized during photo-generated carrier migration.

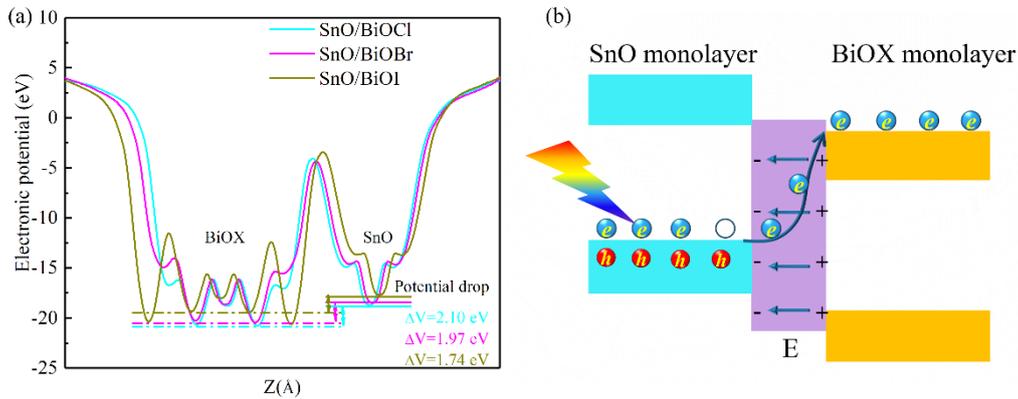

Fig. 5 The calculated electronic potential (a) and the diagrammatic view illustrating the photoexcitation process under electric built-in field (b) of SnO/BiOX bilayer.

In addition to the large difference of effective masses, the electrostatic potential is also calculated to analyze the interfacial characteristics of bilayers, as shown in Fig. 5. The corresponding potential drops are 2.10, 1.97 and 1.74 eV, respectively, indicating the electrons being driven from BiOX to SnO (see Fig. 1(b)). Net charge accumulation results in forming a built-in electric field at the interface, which hinders the electrons further moving to the SnO layer. When the SnO/BiOX bilayer absorbed photon energy in the range of visible light irradiation, the photo-generated electrons and holes are generated in the conduction band (CB) and valence band (VB), respectively, due to the electrons in the VB pumped to the CB. And because of the existence of built-in electric field, the photo-generated electrons are hard to drop from BiOX to SnO which is beneficial to hinder the recombination of photo-generated carriers. Both the large built-in electric field and the significant difference between $m_e$ and $m_h$ for SnO/BiOX bilayers yield the efficient separation of photo-generated electron-hole pairs. Taking into account the analysis of the partial charge at CBM and VBM, the SnO/BiOX bilayers, excluding the SnO/BiOI bilayer, are the most promising candidate for the water splitting due to the rich effective photo-generated carries a.

**CONCLUSION**

In summary, we investigate the electronic structure and optical properties of SnO/BiOX bilayers based on the HSE06 hybrid density functional. The type-II band alignment of SnO/BiOX demonstrate electrons and holes from photo-excitation are mainly derived from different layers and separate in space. The efficient spatial separation of photo-generated carries, the large difference in effective masses and the strong built-in electric field of SnO/BiOX bilayer efficiently hinders the recombination of photo-generated carriers. The optical absorption indicates the absorption onset of SnO/BiOX bilayer extend to the visible light region which enhances the solar light harvesting. Moreover, the band-edge positions relative to the redox potential demonstrate that, exclude the SnO/BiOX bilayer, integrating SnO and BiOX monolayers is a promising strategy to realize the overall water-splitting under visible light irradiation.

**ACKNOWLEDGEMENT**

The work at Beijing Institute of Technology is supported by National Natural Science Foundation of China with Grant Nos. 11572040, 11604011, 11804023 and the China Postdoctoral Science Foundation with Grant No. 2018M641205.

electronic applications. *New Journal of Physics* **2014,** *16* (2), 025005.

**TOC**

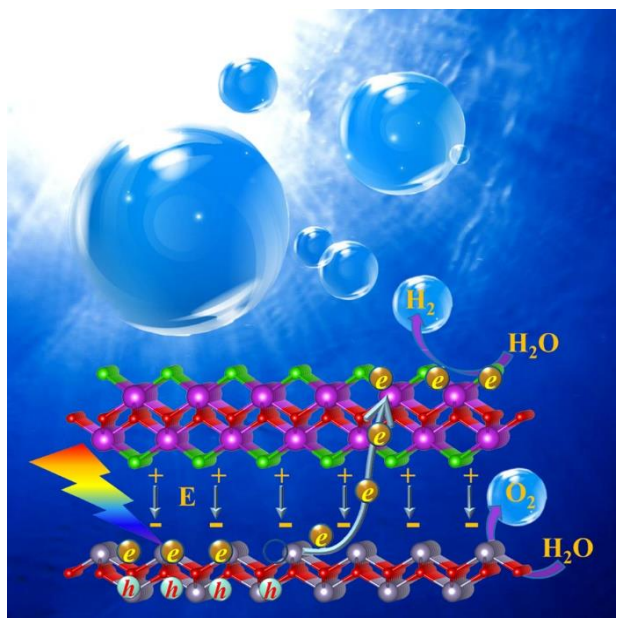